\pdfoutput=1
\documentclass[10pt,twocolumn]{article}
\usepackage[T1]{fontenc}
\usepackage[utf8]{inputenc}
\usepackage[a4paper,top=2.2cm,bottom=2.4cm,left=1.9cm,right=1.9cm,columnsep=0.8cm]{geometry}
\usepackage{newtxtext,newtxmath}
\usepackage{amsmath,cite,url}
\usepackage{graphicx}
\usepackage{float}
\usepackage{booktabs}
\usepackage{placeins}
\usepackage{needspace}
\usepackage{balance}
\usepackage{titlesec}
\titlespacing*{\section}{0pt}{1.1em}{0.5em}
\titlespacing*{\subsection}{0pt}{0.9em}{0.4em}

\newcommand{\secref}[1]{Section~\ref{#1}}
\newcommand{\figref}[1]{Figure~\ref{#1}}
\newcommand{\tabref}[1]{Table~\ref{#1}}
\newcommand{\artifacturl}{\url{https://github.com/JacobLinCool/ChartGenEval}}

\title{\bfseries ChartGenEval: Corruption-Tested Multi-Dimensional
Feedback for Rhythm-Game Chart Generation}

\author{Jhen-Ke Lin\\
  National Yang Ming Chiao Tung University\\
  \texttt{jacob.cs14@nycu.edu.tw}}
\date{}

\newcommand{\titlefigure}{%
  \begin{center}
    \includegraphics[width=0.96\textwidth]{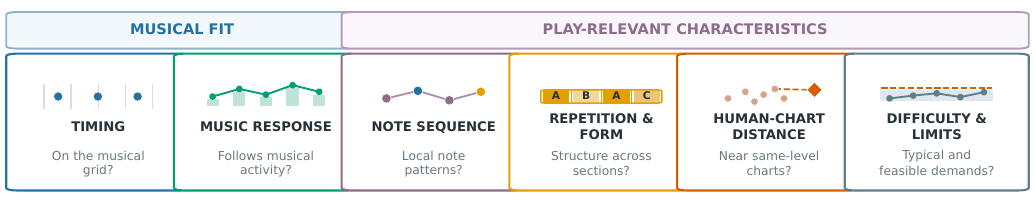}
    \refstepcounter{figure}\label{fig:overview}
    \parbox{0.94\textwidth}{\small
      \textbf{Figure~\thefigure: The six dimensions of ChartGenEval.}
      ChartGenEval organizes chart evaluation into six distinct questions,
      grouped by musical fit and play-relevant characteristics. Each
      dimension retains its own interpretation.}
  \end{center}
}

\makeatletter
\def\@maketitle{%
  \newpage
  \null
  \vskip 2em%
  \begin{center}%
    \let\footnote\thanks
    {\LARGE \@title \par}%
    \vskip 1.5em%
    {\large
      \lineskip .5em%
      \begin{tabular}[t]{c}%
        \@author
      \end{tabular}\par}%
    \vskip .7em%
    {\large \@date}%
  \end{center}%
  \par
  \vskip .6em%
  \titlefigure
  \vskip .8em%
}
\makeatother

\begin{document}
\maketitle

\begin{abstract}
A generated rhythm-game chart need not reproduce one official note
sequence: many note choices can fit the same song and difficulty.
Reference-note agreement therefore measures reconstruction, not the full
design problem. We introduce ChartGenEval, a six-question evaluation
framework with an automatic, corruption-tested core. It leaves note choice
open while anchoring timing to the song: the matched official chart supplies
only its authored timing map, never target notes.

We test each core output with dose-controlled failures rather than assume
that a familiar statistic measures chart quality. Across 80 held-out song
groups, seven output axes satisfy prespecified sensitivity and invariance
criteria in nine nonredundant tests. Complementary stress tests on the
40-song development panel expose two broader lessons. A chart-wide phase
estimate recovers injected shifts of 15, 30, and 60\,ms while chart-only
outputs remain essentially unchanged. Common-pattern rewriting lowers
mean language-model perplexity by 37\%, and loop collapse raises mean
self-similarity by 62\%.
ChartGenEval therefore reports separate, role-specific signals instead of one
proxy or total score. This profile provides automatic feedback for comparing
and iterating generators; selected outputs are candidate optimization targets
or constraints after task-specific stress testing.
\end{abstract}

\section{Introduction}\label{sec:intro}

Chart evaluation is difficult because charting is a design task: multiple
timed, typed note sequences can fit the same song and difficulty. Automatic
generation has progressed from onset detection to conditional sequence
models~\cite{donahue2017ddc,halina2021taikonation,takada2023genelive,
hanzen2025tcp}, but evaluation must still distinguish valid alternatives from
errors.

Agreement with one official chart is therefore useful but incomplete. A
placement F1 score measures reconstruction: a valid alternative loses credit
in the same way as an error. Other automatic proxies answer narrower
questions. Language-model
(LM) likelihood measures model fit, while diversity and self-similarity
describe generated material. None alone tells a developer which chart
property changed or how to improve the generator. Fast model iteration needs
automatic feedback that keeps these questions separate.

Our key observation is that freedom of note choice does not remove external
constraints. A chart may choose \emph{what} notes to place, but it cannot
choose the song's musical time. We therefore use the matched official chart
only as an authored map of bars, meter, and tempo. Its note placements never
become targets. This separation preserves alternative designs while exposing
chart-wide timing shifts that interval- and type-based statistics cannot see.

ChartGenEval organizes evaluation into the six questions in
\figref{fig:overview}. Its corruption-tested core covers timing, local note
sequence, local repetition, and difficulty or playability limits. Each output
has a role: a directional signal can guide optimization, a feasibility check
can impose a constraint, and a descriptive value can monitor model behavior.
The profile does not collapse these roles into one score.

We test the core outputs by injecting known failures at increasing strengths.
A useful output should react to its target failure and remain unchanged under
predeclared invariance controls. Seven output axes meet these criteria in nine
nonredundant tests on 80 held-out song groups. Complementary development
stress tests yield two broader lessons. First, timing must be anchored to the
song: our phase estimate recovers 15--60\,ms chart-wide shifts that chart-only
statistics miss. Second, a plausible proxy can reward the wrong behavior:
common-pattern rewriting improves perplexity, while loop collapse increases
self-similarity.

Our contributions are:
\begin{enumerate}
\item \textbf{Automatic, dimension-specific feedback.} ChartGenEval
  separates evaluation questions and output roles, allowing generators to be
  compared and revised without treating one reference note sequence as the
  only valid answer.
\item \textbf{Corruption-tested signals and reusable diagnostic lessons.}
  Nine nonredundant held-out tests support seven core output axes.
  Complementary development tests identify a translation blind spot in
  chart-only timing statistics and wrong-way incentives in perplexity and
  self-similarity.
\end{enumerate}

\section{Related Work}\label{sec:related}

\textbf{Reference-chart agreement measures reconstruction against one
authored sequence.} DDC and Gen\'eLive! report F-scores
for predicted placements, while TCP adds onset-plus-type F1 and local pattern
recall~\cite{donahue2017ddc,takada2023genelive,hanzen2025tcp}.
TaikoNation compares binary timing frames and pattern
spaces~\cite{halina2021taikonation}. These measures remain valuable for
reconstruction, but they treat one authored sequence as the answer.

\textbf{Unpaired statistics and human studies answer complementary
questions.} DDC uses LM perplexity
and token accuracy to measure held-out model fit~\cite{donahue2017ddc}.
Symbolic-music and text-generation work adds diversity, distribution,
self-similarity, and repetition statistics~\cite{li2016diversity,
zhu2018texygen,yang2020evaluation,dong2020muspy,wu2020jazz}. Human and expert
studies instead evaluate system-level experience~\cite{gover2022mute,
takada2023genelive,wang2024mania}. ChartGenEval complements these approaches
by testing whether automatic outputs respond to specified chart failures.

\textbf{Metric stress tests show that a plausible statistic can miss a
musical failure.} SCHmUBERT constructs
non-musical piano rolls that match framewise self-similarity
statistics~\cite{plasser2023schmubert}. Fr\'echet Music Distance is tested with
pitch and velocity perturbations, but remains a set-level embedding
distance~\cite{retkowski2025fmd}. STRUM reports reference-onset F1 after
a per-song global-offset search~\cite{opria2026strum}; ChartGenEval
reports the offset itself without using reference notes as targets. Its
controlled corruptions test both target response and invariance, turning
metric choice into an empirical question.

\section{Methods}\label{sec:methods}

\subsection{Evaluation questions and core outputs}\label{sec:suite}

ChartGenEval treats the six dimensions in \figref{fig:overview} as separate
questions. The main analysis focuses on seven outputs for which we have
held-out controlled-corruption evidence (\tabref{tab:suite}). Timing uses
three readings because dense noise, rare outliers, and a shared offset are
different failures. The other outputs cover local transitions, local
repetition, requested note rate, and density spikes.

\begin{table*}[t]
\centering
\footnotesize
\setlength{\tabcolsep}{6pt}
\renewcommand{\arraystretch}{1.15}
\begin{tabular}{@{}p{0.15\textwidth}p{0.50\textwidth}p{0.27\textwidth}@{}}
\toprule
Evaluation question & Automatic reading & Present evidence \\
\midrule
Timing & Timing clean rate, timing-error p99, and whole-chart grid-phase offset & C1, C1s, and C2 held-out tests \\
Note sequence & Transition familiarity: a one-sided score derived from the rate of unseen interval--type trigrams & C3 held-out test \\
Repetition and form & Same-course typicality of repeated versus unique 4-grams & C4, C5, and C8 test local repetition; long-form readings remain exploratory \\
Response to music & Bar-scale energy response and onset support & Exploratory development analysis; Appendix~\ref{app:visuals} \\
Human-chart distance & Distance from the same-course training distribution & Exploratory development analysis; Appendix~\ref{app:visuals} \\
Difficulty and limits & Same-course note-rate typicality and a local density-spike limit & C6 and C7 held-out tests \\
\bottomrule
\end{tabular}
\caption{Evidence differs across the six questions: seven core outputs have
held-out tests, while the response-to-music, human-distance, and long-form
readings remain exploratory.}
\label{tab:suite}
\end{table*}

All outputs are automatic, but automation does not make them
interchangeable. Grid-phase error and timing tails have a clear direction;
density spikes provide a course-relative density constraint; transition familiarity and
typicality are safer as soft targets or monitoring signals. Raw LM loss is
retained only as a description of model fit and as a stress-test baseline.
In particular, an unseen transition is unfamiliar to the training corpus,
not necessarily invalid.

\subsection{Same-difficulty calibration}\label{sec:layers}

Several chart properties have no universal ``more is better'' direction. A
chart can be too sparse or too dense, and too little or too much repetition
can both be undesirable. We therefore measure departure from the middle 80\%
of same-course human charts. For a feature value $x$, let $l$ and $u$ be the
10th and 90th training percentiles and
$h=\max((u-l)/2,10^{-9})$. The band score is

\begin{equation}
\begin{aligned}
d(x)&=
\begin{cases}
(l-x)/h, & x<l,\\
0,       & l\leq x\leq u,\\
(x-u)/h, & x>u,
\end{cases}\\
s(x)&=\frac{1}{1+d(x)^2}.
\end{aligned}
\label{eq:band}
\end{equation}
Thus $s(x)=1$ inside the band and decreases symmetrically outside it. This is
a difficulty-conditioned typicality reading, not a verdict on creativity or
overall quality. One-sided limits, such as the density-spike check, remain
separate because a severe local overload should not be canceled by another
high score.

The three less familiar outputs use simple transformations. If $r$ is the
rate of interval--type trigrams unseen in same-course training charts,
transition familiarity is $\exp(-r/0.08)$. For $M$ 4-gram windows with $U$
unique windows, the raw repetition rate is $(M-U)/M$ before band calibration.
The density-spike score is $\exp(-e/1.5)$, where $e$ is the amount by which a
chart's 95th-percentile local density jump exceeds the same-course limit.

\subsection{Data and timing anchors}\label{sec:data}

We separated design from confirmation. The seven core outputs and their
corruptions were designed on 40 development song groups with 170 charts, then
tested on 80 held-out groups containing 333 charts. The calibration set
contains 3{,}880 course charts from 813 separate song groups. Songs, rather
than courses or corruption replicates, are the inferential units.
Appendix~\ref{app:protocol} gives the corpus construction, completeness rules,
and full analysis protocol.

Timing requires a reference outside the generated event sequence. Our primary
grid comes from the matched official file's authored bar timestamps, meter,
and tempo. We never use its note times or note types as targets, and we never
accept the grid declared by the generated chart. The first choice preserves
freedom of note design; the second prevents a generator from redefining the
clock against which it is evaluated. An audio-estimated alternative is
reported in Appendix~\ref{app:protocol}.

\subsection{Measuring timing against the song}\label{sec:timingmethod}

A common shift $\delta$ leaves every inter-note interval unchanged:
\begin{equation}
(t_{i+1}+\delta)-(t_i+\delta)=t_{i+1}-t_i.
\label{eq:shift-invariance}
\end{equation}
The note types are also unchanged. Chart-only measurements built from
these intervals and types therefore cannot identify the shift. This
translation symmetry explains the blind spot before any experiment is run.

Local timing still needs both a common-error summary and a tail summary. We
treat a note within 6\,ms of the authored grid as locally aligned, at the scale
of published temporal discrimination for short
intervals~\cite{friberg1995jnd}. Timing clean rate summarizes common errors;
the 99th percentile of absolute error exposes rare large errors that a mean
can hide. Unmatched generated notes remain in the denominator, so timing clean
rate must be read with note rate. Full matching rules and the 6/12/18\,ms
diagnostic ranges appear in Appendix~\ref{app:protocol}.

\begin{figure*}[t]
\centering
\includegraphics[width=0.94\textwidth]{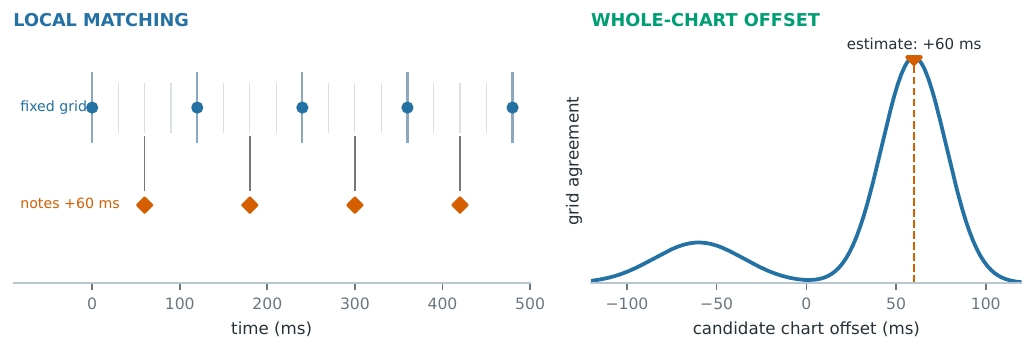}
\caption{Why timing needs a whole-chart estimate. \textbf{Left:}
nearest-grid matching can pair shifted notes with different subdivision
points and report small local errors. \textbf{Right:} a schematic agreement
curve over candidate whole-chart offsets; the curve peaks at the injected
$+60$\,ms shift. Local plausibility does not imply correct song alignment.}
\label{fig:timing}
\end{figure*}

Nearest-grid matching can miss a constant shift (\figref{fig:timing}).
On a dense grid, each shifted note may simply be paired with another
nearby grid point. We therefore estimate one offset for the whole chart.
Let $G_k$ be the fixed bar, beat, and eighth-note grids, and let
$d(x,G_k)=\min_{g\in G_k}|x-g|$. For note times $t_i$, we score a
candidate shift by
\begin{equation}
\begin{aligned}
S(\delta)
 &= \sum_k w_k\,\frac{1}{n}\sum_i
 \left[1-\frac{d(t_i-\delta,G_k)}{\tau}\right]_+,\\
\hat{\delta}
 &= \operatorname*{arg\,max}_{\delta\in[-250,250]\,\mathrm{ms}}
 S(\delta),
\end{aligned}
\label{eq:phase-offset}
\end{equation}
where $[z]_+=\max(z,0)$, $\tau=12$\,ms, and the bar, beat, and
eighth-note weights are 0.2, 0.3, and 0.5. We search in 0.5\,ms steps.
Combining three grid levels reduces periodic ties. The resulting
$\hat\delta$ is a chart-level diagnostic: its magnitude says how far the
chart must move to best align with the fixed musical-time framework. BPM alone
provides a period but not a phase, so it cannot replace this anchor.

\subsection{Controlled corruptions and acceptance rules}
\label{sec:probes}

Starting from human charts, each corruption strengthens one operational
failure. Other outputs may also move, so the tests establish target
sensitivity rather than exclusive specificity. The target output and expected
direction were fixed before the held-out run; all transformations appear in
Appendix~\ref{app:corruptions}.

After orienting outputs so larger is better, support requires a negative
dose-rank association, a negative strongest-dose contrast against the matched
control, and every relevant invariance check to pass. We use song-cluster
bootstrap intervals adjusted across ten prespecified rows. Full decision rules
and sample accounting appear in Appendix~\ref{app:protocol}.

\section{Results}\label{sec:results}

\subsection{Core outputs respond to target failures}
\label{sec:sensitivity}

Every core output has the expected negative dose-rank association under its
target failures (\tabref{tab:confirmation}). The result covers nine nonredundant
corruption--measurement pairs but seven output axes because the same 4-gram
axis detects three distinct edits. All applicable invariance controls pass.
The important result is not that every corruption needs a new metric. A small
set of interpretable outputs can cover different failure mechanisms when each
output is tested against a concrete claim.

\begin{table*}[t]
\centering
\scriptsize
\setlength{\tabcolsep}{3.5pt}
\begin{tabular}{@{}p{0.20\textwidth}p{0.21\textwidth}p{0.25\textwidth}p{0.26\textwidth}@{}}
\toprule
Targeted failure & Core output & Dose trend [99.5\% CI] &
Strongest-dose change [99.5\% CI] \\
\midrule
Dense timing jitter & timing clean rate & $-.744\;[-.815,-.667]$ & $-.502\;[-.558,-.446]$ \\
Sparse 60\,ms timing errors & timing-error p99 & $-.234\;[-.344,-.134]$ & $-.306\;[-.508,-.144]$\,ms \\
Whole-chart shift & grid-phase offset & $-.916\;[-.981,-.826]$ & $-55.58\;[-61.43,-49.83]$\,ms \\
Note-type shuffle & transition familiarity & $-.800\;[-.856,-.737]$ & $-.231\;[-.262,-.199]$ \\
Loop collapse & 4-gram typicality & $-.445\;[-.573,-.312]$ & $-.165\;[-.226,-.108]$ \\
Common-pattern rewrite & 4-gram typicality & $-.358\;[-.459,-.258]$ & $-.131\;[-.189,-.078]$ \\
Note-rate scaling & note-rate typicality & $-.693\;[-.784,-.579]$ & $-.582\;[-.672,-.483]$ \\
Local burst insertion & density-spike limit & $-.940\;[-.950,-.930]$ & $-.619\;[-.644,-.593]$ \\
Bar-order shuffle & 4-gram typicality & $-.589\;[-.728,-.435]$ & $-.224\;[-.293,-.156]$ \\
\bottomrule
\end{tabular}
\caption{Held-out results on 80 song groups. Outputs are oriented so larger is
better; a useful response is therefore negative as a failure becomes stronger.
Every row meets both criteria, and all applicable controls pass. Intervals are
family-wise 99.5\% song-cluster bootstrap intervals.
The two originally named 4-gram variety and repetition scores are algebraic
mirrors, so they appear once as one effective axis.}
\label{tab:confirmation}
\end{table*}

Sparse errors show why a tail statistic is needed when most notes remain
correct. Complete sample counts and exclusions appear in
Appendix~\ref{app:protocol}.

\subsection{External time detects chart--audio shifts}
\label{sec:anchorresults}

Translation symmetry predicts that a chart can look internally plausible yet
be late relative to the song. The 40-song development experiment confirms
this mechanism. After shifting every note by 15, 30, or 60\,ms, no primary
chart-only output changes by more than 0.03 human standard deviations. The
phase estimator's median matches each injected shift. With either the authored
or audio-estimated grid, at least 169 of 170 charts are within 3\,ms of the
injected value.

Local matching and global phase answer different questions. A dense grid can
re-pair a shifted note to another subdivision and still report a small local
error. The whole-chart estimate preserves the shared offset instead. The same
principle explains sparse failures: the mean remains dominated by correct
notes, whereas timing-error p99 exposes the tail. Timing evaluation therefore
needs both a song-side anchor and summaries at more than one scale.

\subsection{Common proxies can reward the wrong behavior}
\label{sec:reductio}

\begin{figure}[H]
\centering
\includegraphics[width=\columnwidth]{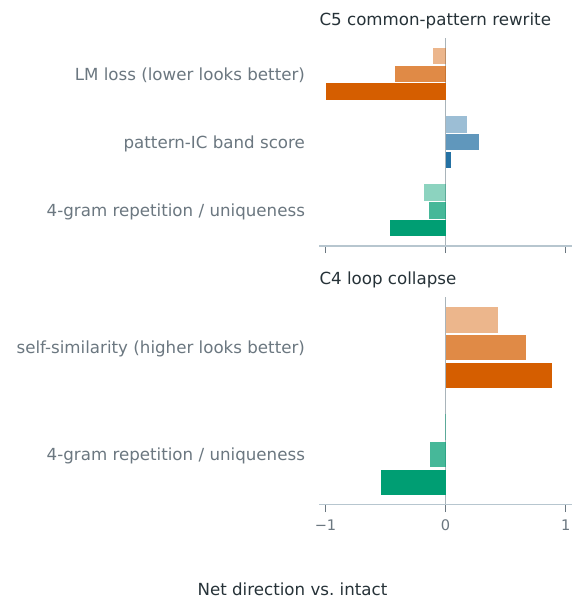}
\caption{Wrong-way incentives on the development panel. Common-pattern
rewriting improves LM loss and its band score; loop collapse raises
self-similarity. In both cases, 4-gram typicality falls. Bars show net direction
from the intact chart; darker bars mark stronger corruption.}
\label{fig:complementarity}
\end{figure}

Two familiar proxies move in the wrong direction under targeted failures
(\figref{fig:complementarity}). LM perplexity rewards common-pattern
rewriting: on the development panel, mean perplexity falls by 37\%, from 9.44
to 5.98. A lower-is-better reading therefore prefers the corrupted chart,
opposite to the intended property~\cite{theis2016note}. The reversal remains
with a separately fitted scoring LM, while 4-gram typicality moves away from
the human range. Secondary checks appear in Appendix~\ref{app:visuals}.

If self-similarity is maximized as a quality objective, it rewards loop
collapse. Repeating the dominant 4-gram raises mean self-similarity by 62\%
even though the note-type stream has collapsed into a four-event
cycle~\cite{wu2020jazz}.
4-gram typicality moves in the opposite direction. The practical rule is to
stress-test a candidate reward against the failure it is meant to prevent.

Automatic profiles shorten model-development feedback by identifying the
failure mechanism that changed. Phase error and timing tails can guide
minimization; density spikes provide a course-relative constraint; familiarity
and typicality are soft targets or monitors. Optimization may use selected
signals after task-specific stress testing, but no single number should become
the reward.

\FloatBarrier

\section{Discussion}\label{sec:discussion}

Typicality also needs context. A corruption often pushes a chart outside the
same-course human band, but the reverse implication is invalid: an unusual
chart may be deliberate and playable. Separate timing and feasibility checks
help distinguish novelty from a concrete failure. When signals disagree, the
disagreement is information rather than an error to hide with an average.

The descriptive public-system profiles in Appendix~\ref{sec:systemresults}
illustrate two reusable reading habits. Similar clean-note fractions can hide
different timing tails and global offsets, and a generator can match one
difficulty while overshooting another. These examples are not additional
metric validation; they show why diagnosis should follow failure mechanisms
rather than one headline score.

\section{Limitations}\label{sec:limitations}

Controlled corruptions establish response to defined failures, not a universal
ranking of chart quality. The present set does not cover every error, such as
wrong-song pairing or removal of musically important notes. The response-to-
music, long-form, and human-distance extensions were designed on the
development panel and remain exploratory.

The human bands come from one Taiko corpus. Timing also requires an external
grid; without an authored map or a reliable audio estimate, the timing outputs
are unavailable. Finally, timing clean rate has no reference-note recall term,
so deleting off-grid notes can improve it. This is why the profile reads clean
rate together with note rate and keeps feasibility checks separate.

\FloatBarrier

\section{Conclusion}\label{sec:conclusion}

ChartGenEval turns chart evaluation into separate, testable feedback signals.
Seven output axes pass nine held-out controlled-corruption tests. On the
development panel, the song-anchored phase estimate recovers injected shifts
invisible to chart-only statistics. Those development tests also show that
likelihood and self-similarity can move in the wrong direction under targeted
failures, so neither should be used alone as a quality score. The broader lesson is simple:
automatic evaluation is most useful when each signal names a failure, survives
a targeted stress test, and keeps its role visible to the model developer.

\section*{Ethics Statement}
The study uses community-transcribed charts and associated audio
gathered from online sources as described in \secref{sec:data}; these
potentially copyrighted materials are not redistributed. The released
materials exclude source charts, audio, personal data, and generators.
Automated evaluation can lower the cost of content production, so high-volume
deployment still requires platform policy and moderation.

\section*{AI Usage Statement}
AI assistants were used for experiment orchestration, drafting, and
editing under author direction; the author checked all reported values.

\appendix

\section{Corpus Coverage and Held-Out Evaluation Protocol}\label{app:protocol}

\subsection{Corpus coverage}

\begin{figure*}[t]
\centering
\includegraphics[width=0.98\textwidth]{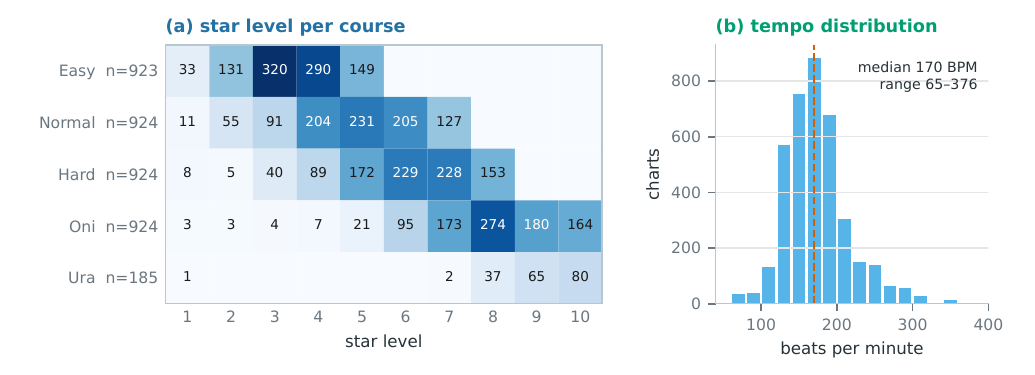}
\caption{Course difficulty and tempo coverage of the human reference corpus
that defines the calibration bands. \textbf{(a)} Star-level counts per course
across the 3{,}880 training charts: each course spans a broad, shifting
difficulty range. \textbf{(b)} Tempo histogram (median 170\,BPM, range
65--376).}
\label{fig:coverage}
\end{figure*}

The calibration corpus is summarized by course, star level, and tempo.
\figref{fig:coverage} shows its difficulty and tempo coverage.

\subsection{Corpus and timing details}

We use Taiko-style chart files and associated audio gathered from community
resources available online. The training split contains 924 source rows that
resolve to 813 unique song groups and 3{,}880 course charts. The test split
contains 140 source rows that resolve to 120 song groups. The underlying charts
and audio may be copyrighted, so they are not redistributed.

The primary timing grid uses authored per-bar timestamps, meter, and tempo.
For local timing diagnostics, every generated note is classified as within
6\,ms, in the 6--12, 12--18, or above-18\,ms range, or unmatched. We also
compare adjacent intervals with tolerance
$\max(6\,\mathrm{ms},2.5\%\times\mathrm{IOI})$. An audio-downbeat estimate
provides a development-stage alternative: its period is within 2\% of the
metadata period on 39 of 40 development songs and differs by 2.39\% on the
remaining song. Held-out absolute offsets use the authored grid because it
fixes bar phase directly.

\subsection{Panel construction and integrity}

Source rows are canonicalized by \texttt{group\_id} and audio SHA-256;
inconsistent aliases are rejected before panel indexing. The test split
contains 120 canonical song groups in stored order. Groups 0--39 form
the development panel, and groups 40--119 form the held-out panel.
Before inspecting held-out results, we fixed the dataset revision,
corruptions and doses, target measurements, controls, thresholds,
calibration, and analysis plan.

\subsection{Replicates, completeness, and inferential units}

Every nonzero corruption cell and control condition uses five fixed base
seeds. Replicates are averaged within chart and dose. A course is
complete only when its intact value, every target
dose and replicate, matched control, and required checks are available. A
song enters a test with at least one complete course, and complete
courses are averaged within song. Songs, rather than charts, courses, or
replicates, are the inferential units for both response statistics.
Deterministic no-op and missing cells do not enter the analysis.

\subsection{Control checks and numerical invariance}

Identity re-evaluation applies to every retained measurement. A joint
+1-second shift of notes, authored grid, and duration tests invariance to
the time origin. A Don--Ka color bijection applies to timing and
density, where color is irrelevant; it is not a grammar or form control.
The 60\,ms C2 translation is the matched control for chart-only density, grammar,
and structure measurements. A required control failure prevents support
for the affected target response. Each measurement uses a specified
equivalence tolerance and timestamp-rounding rule.
Chart-only interval arithmetic uses integer microseconds, three orders
of magnitude finer than the smallest reported tolerance.

\subsection{Intervals and decision rule}

After orienting outputs so that larger is better, the first statistic is
the within-chart Spearman correlation between the four severities
(intact plus three doses) and the measurement; an all-constant response
has zero sensitivity. The second statistic is the oriented
maximum-dose value minus the matched control. Point estimates average
song-level values. We use 10{,}000 song-cluster bootstrap draws:
secondary analyses report nominal 95\% intervals, and the ten primary
rows use Bonferroni-adjusted 99.5\% intervals.
C6 assigns severities $[0,0.5,0.5,1]$ to the intact, $0.5\times$,
$1.5\times$, and $2\times$ conditions, so equal-magnitude low and high
departures are tied; tied severities receive average ranks.
A response is supported when at least 72 of 80 songs are evaluable, all
controls pass, and both adjusted upper bounds are below zero. A
nonnegative adjusted lower bound contradicts the response; other cases
are inconclusive. The two C5 output rows were required to meet the rule
together, although their algebraic equivalence leaves one effective
axis. The C5 LM uses the full stored-order training split, trigram order
3, and add-$\alpha=0.05$.

\subsection{Sample counts and exclusions}

The held-out execution produced 50{,}283 observation records from 333
charts. C1s retained 332 complete courses and C8 retained 330; every
other effective pair retained 333. All 32 controls passed with maximum
observed absolute difference zero. Nineteen deterministic no-op cells
were excluded.

\FloatBarrier

\section{Controlled-Corruption Specifications}\label{app:corruptions}

\begin{table*}[t]
\centering
\footnotesize
\setlength{\tabcolsep}{4pt}
\begin{tabular}{@{}p{0.30\textwidth}p{0.18\textwidth}p{0.46\textwidth}@{}}
\toprule
Controlled corruption (strength) & Operational change & Prespecified response \\
\midrule
C1 timing jitter ($\pm$10/20/30\,ms) & all note times & timing clean rate $\downarrow$ \\
C1s sparse timing errors (0.5/1/2\%, $\pm$60\,ms) & a few note times & timing-error p99 $\uparrow$ \\
C2 whole-chart shift (+15/30/60\,ms) & chart--song alignment & grid-phase offset $\uparrow$ \\
C3 note-type shuffle ($p{=}0.2/0.4/0.8$) & local transitions & transition familiarity $\downarrow$ \\
C4 loop collapse ($p{=}0.3/0.6/1$) & local repetition & 4-gram typicality $\downarrow$ \\
C5 common-pattern rewrite ($p{=}0.3/0.6/1$) & pattern variety & 4-gram typicality $\downarrow$ \\
C6 note-rate scaling ($\times$0.5/1.5/2) & difficulty fit & note-rate typicality $\downarrow$ \\
C7 burst insertion (2/4/8 clusters) & local overload & density-spike limit $\downarrow$ \\
C8 bar-order shuffle ($p{=}0.3/0.6/1$) & phrase order & 4-gram typicality $\downarrow$ \\
\bottomrule
\end{tabular}
\caption{Controlled corruptions, strengths, and target readings. The edits
target operational failures rather than claim to exhaust chart quality.
C3--C5 preserve note times; C4 and C6 preserve or control density.}
\label{tab:probes}
\end{table*}

\begin{figure*}[t]
\centering
\includegraphics[width=0.98\textwidth]{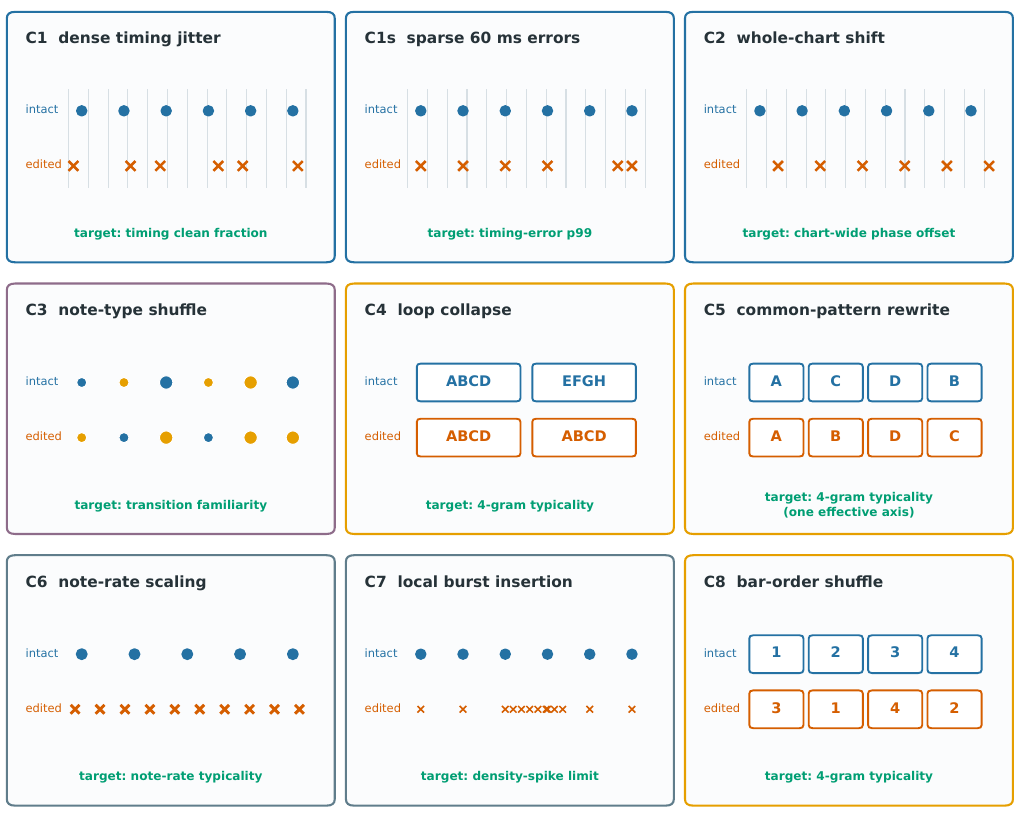}
\caption{Controlled-corruption atlas. Blue marks the intact chart and
orange marks edited events. Each panel shows the property targeted by one
corruption and names the target measurement. C1--C2 target timing, C3
targets transition familiarity, C4--C5 and C8 target repetition or
form, and C6--C7 target difficulty fit and local load. These edits do
not exhaust chart quality.}
\label{fig:corruptionatlas}
\end{figure*}

\figref{fig:corruptionatlas} visualizes the operational change made by
each corruption and its prespecified target measurement.

The 170-chart development matrix uses one realization per nonzero cell,
producing 4{,}760 variants including the originals. C1s moves
$\max(1,\operatorname{round}(np))$ of a chart's $n$ notes by
$\pm60$\,ms. C3--C5 preserve note times, while C4 and C6 preserve or
control density. Raw features and their derived scores are evaluated on
the same variants; equivalent quantities are counted once. For C5, one
LM chooses common-pattern rewrites and a separately fitted LM scores
them. The perplexity reversal remains (9.44 $\to$ 6.04, $-36\%$).

\FloatBarrier

\section{Metric Guide and Extended Results}\label{app:visuals}

\subsection{Exploratory measurements}

Five additional components were selected or revised on the 40-song
development panel. They remain useful hypotheses, but the main paper does not
treat them as held-out-confirmed targets. \emph{Reciprocity} compares two-bar
phrases and rewards material that returns after an intervening phrase without
rewarding adjacent copying. \emph{Stagnation--alienation} marks active bars in
long repeated runs or without a related bar within a 32-bar neighborhood.
\emph{Density--energy response} is the rank correlation between bar-level note
density and mel energy. \emph{Run-head onset support} asks whether the first
note of a rhythmic group lies near strong spectral flux. \emph{Human-chart
distance} measures the mean distance to the 20 nearest same-course training
charts in a standardized 32-feature representation.

Their development responses clarify their scope. Reciprocity reacts to loop
collapse and bar shuffling; stagnation--alienation reacts most strongly to
loop collapse. Run-head support detects whole-chart shifts, while
density--energy response detects shuffled bars but is weak for short bursts.
Human-chart distance detects many distributional departures but does not say
whether an unusual chart is poor or creative. These outputs are therefore
reported as exploratory monitors rather than core rewards.

\subsection{Proxy diagnostics}

The common-pattern rewrite is not an artifact of using the same LM for editing
and scoring. When one LM selects the rewrite and a separately fitted LM
scores it, perplexity still falls from 9.44 to 6.04 ($-36\%$). A symmetric
band transform of LM loss also fails to identify the maximum rewrite dose: its
held-out change is $+0.021$ with a nominal 95\% interval of
$[-0.034,0.077]$. The 4-gram axis supplies the missing reading. Under loop
collapse, set-level Self-BLEU rises from 0.845 to 0.898, also recording the
loss of diversity.

\begin{figure*}[t]
\centering
\includegraphics[width=0.98\textwidth]{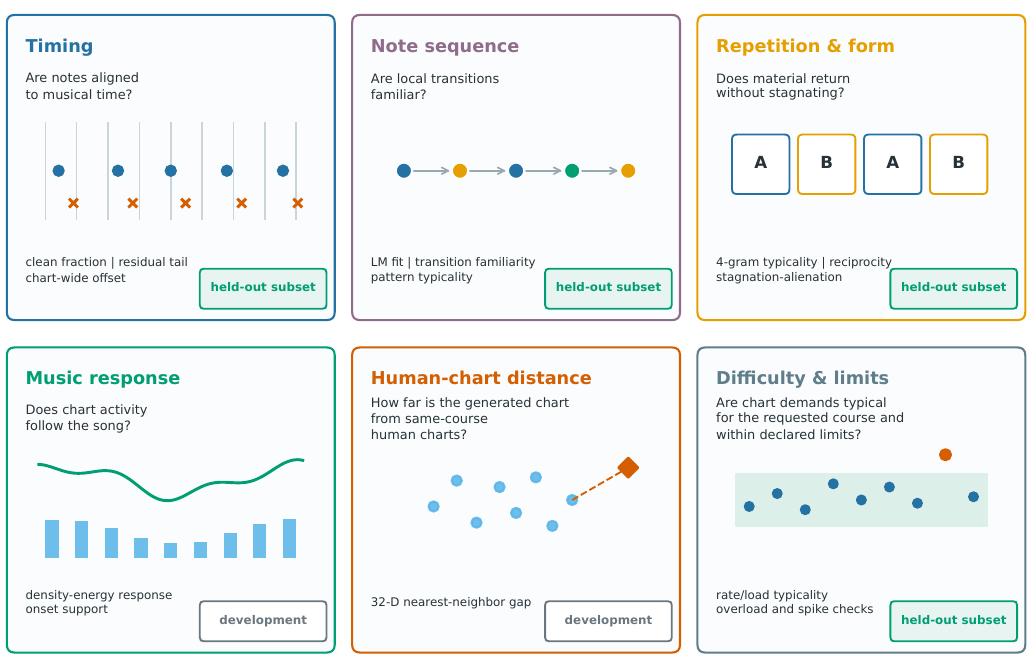}
\caption{Metric-to-construct guide. Each panel states an evaluation
question, illustrates the relevant signal, lists representative outputs,
and marks the strongest evidence. ``Held-out subset'' means that only
the named outputs in \tabref{tab:suite} have held-out corruption
evidence. Typicality describes departure from same-course human ranges.}
\label{fig:metricguide}
\end{figure*}

\figref{fig:metricguide} connects each construct to its required inputs,
representative outputs, and strongest present evidence. The evidence
labels distinguish outputs with held-out corruption tests from components
supported by development analyses.

\begin{figure*}[t]
\centering
\includegraphics[width=0.94\textwidth]{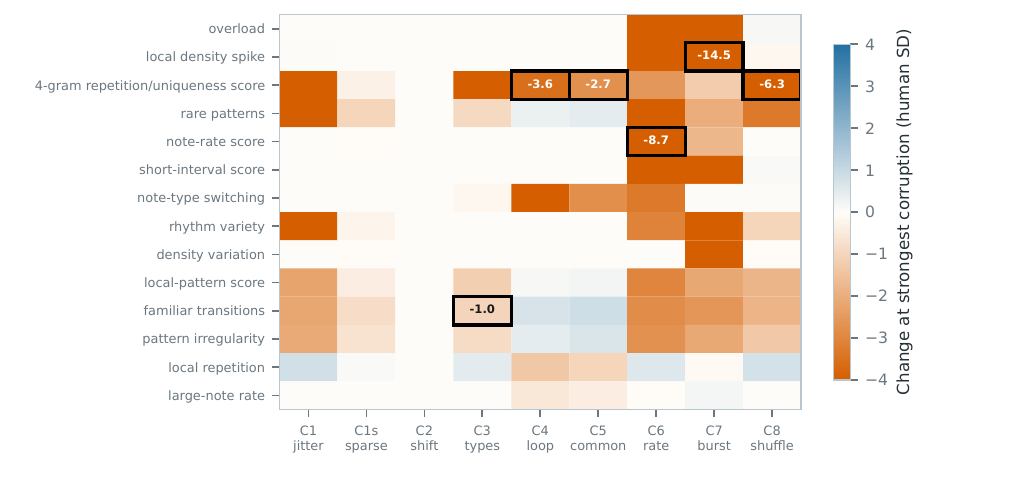}
\caption{How 14 nonredundant calibrated chart-only scores respond to
each corruption at maximum strength on the development panel. Cells
cover band-typicality scores and one-sided checks. Red means the score
falls and blue that it rises; values are in standard deviations of the
human reference charts, clipped at $\pm4$. Black outlines mark the
target rows selected on the development panel; they do not mark
statistical significance. The matrix shows both target sensitivity and
cross-family coupling.}
\label{fig:responsemap}
\end{figure*}

\figref{fig:responsemap} shows that targeted corruptions can affect
multiple score families, motivating interpretation as a profile rather
than as isolated interchangeable measurements.

\FloatBarrier

\subsection{Public-system profiles}\label{sec:systemresults}

We apply the profile to five public generators on the 40 development songs.
Mapperatorinator v32~\cite{mapperatorinator} and
TaikoNation~\cite{halina2021taikonation} produce Taiko note types. DDC's onset
pipeline~\cite{donahue2017ddc}, the public timing configuration of
Gen\'eLive!~\cite{takada2023genelive}, and AutoOsu~\cite{autoosu} enter only
the applicable timing and audio-response readings. We use their public
inference configurations without training or fine-tuning on this corpus.
These profiles illustrate interpretation; they are not held-out metric tests
or a generator leaderboard.

\begin{figure*}[t]
\centering
\includegraphics[width=0.92\textwidth]{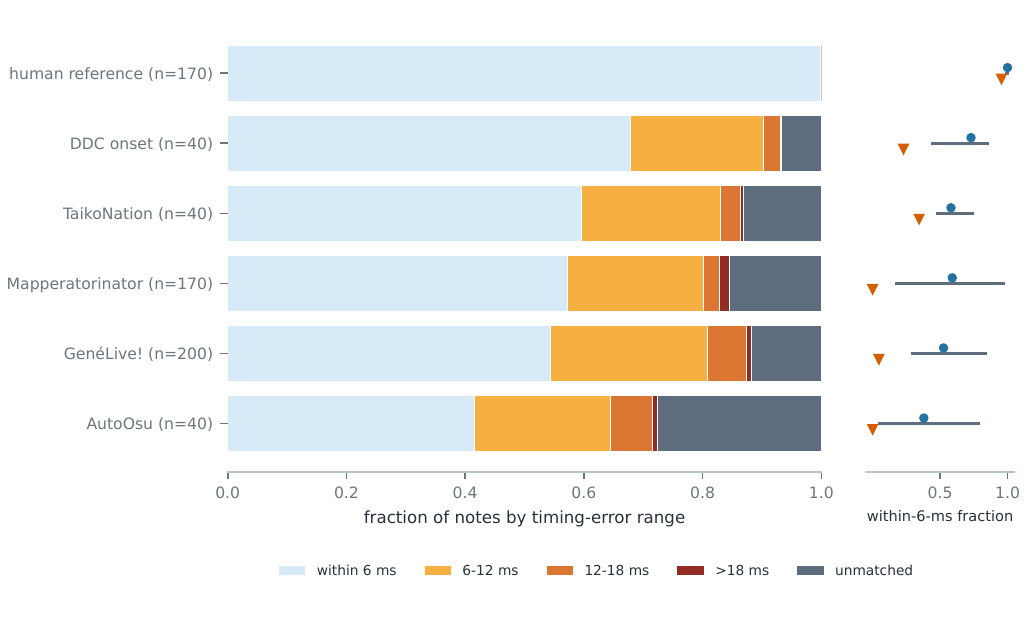}
\caption{Timing profiles on the 40 development songs. The stacked bars separate
locally clean notes from timing tails and unmatched notes; the right strip
shows per-chart variation. Systems with similar mean clean fractions can have
different tails, so one timing average does not identify the failure.}
\label{fig:tiers}
\end{figure*}

\begin{figure*}[t]
\centering
\includegraphics[width=0.92\textwidth]{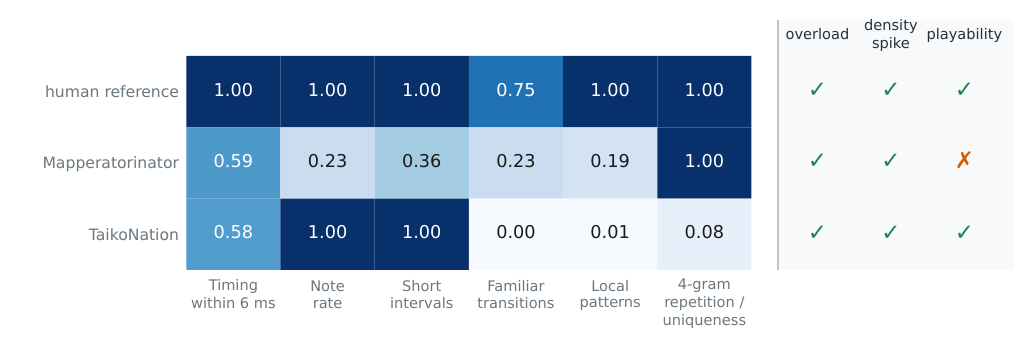}
\caption{Separate development readings for human charts and the two generators
that produce Taiko note types. Columns retain their own meanings; the matrix
is a diagnostic profile, not a total score.}
\label{fig:profiles}
\end{figure*}

The two Taiko generators have similar median clean-note fractions, yet their
relative-interval errors and global phase offsets differ. This is the same
local-versus-global distinction exposed by the shift corruption. Difficulty
conditioning reveals another pattern: Mapperatorinator overshoots human note
rates most on easier courses, while its timing profile changes little by
course. The useful conclusion is a course mismatch, not a general claim that
the system improves at higher difficulty.

\begin{figure*}[t]
\centering
\includegraphics[width=0.96\textwidth]{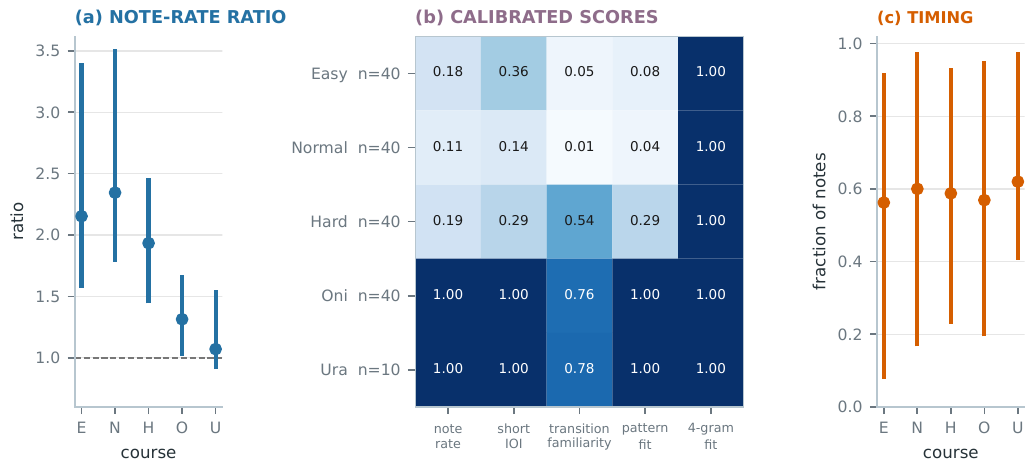}
\caption{Exploratory difficulty profile for Mapperatorinator on paired
development charts. Easy through Oni contain 40 song--course pairs; Ura
contains 10. Dots are medians and vertical lines span the 10th to 90th
percentiles. Generated density approaches the human reference as course
difficulty rises, while timing does not follow the same trend; this
supports a course mismatch rather than general improvement at high
difficulty.}
\label{fig:difficulty}
\end{figure*}

\FloatBarrier

\section{Released Materials}

Code, evaluation records, and plotting scripts are available at
\artifacturl.

\balance
\bibliographystyle{IEEEtran}
\bibliography{references}

\end{document}